\documentclass[twocolumn]{aastex631}
\usepackage{amsmath}
\usepackage{placeins}

\newcommand{\Msun}{M_{\odot}}
\newcommand{\Mbh}{M_{\rm BH}}
\newcommand{\Mdot}{\dot M}
\newcommand{\Ledd}{L_{\rm Edd}}
\newcommand{\rg}{r_{\rm g}}
\newcommand{\Rsph}{R_{\rm sph}}
\newcommand{\Rnu}{R_{\nu}}
\newcommand{\Ruv}{R_{1400}}
\newcommand{\betaUV}{\beta_{\rm UV}}
\newcommand{\angstrom}{\text{\AA}}

\shorttitle{Hidden Supercritical Accretion in GN-z11}
\shortauthors{Mitra \& Santra}

\begin{document}

\title{A Standard Ultraviolet Continuum Can Hide Supercritical Accretion in GN-z11}

\author{Samik Mitra}
\affiliation{International Centre for Theoretical Sciences, Tata Institute of Fundamental Research, Bengaluru 560089, India}
\email{samik.mitra@icts.res.in}

\author{Ramananda Santra}
\affiliation{International Centre for Theoretical Sciences, Tata Institute of Fundamental Research, Bengaluru 560089, India}
\email{ramananda.santra@icts.res.in}
\correspondingauthor{Samik Mitra}

\begin{abstract}
The ultraviolet continuum from a high-redshift accreting black hole is routinely used to infer its mass. GN-z11 offers a sharp test of whether that inference is secure. A thin-disk fit to its continuum gives an Eddington mass of $M_{\rm E}=1.12\times10^{7}\,\Msun$, an order of magnitude above the $\log(\Mbh/\Msun)=6.2\pm0.3$ inferred from broad N~\textsc{iv}. We show that this apparent tension is not imposed by the ultraviolet slope alone. We derive a closed-form criterion comparing the radius where observed photons are produced with the radius where a supercritical flow departs from efficient solution. For GN-z11, the criterion places the supercritical transition inside the ultraviolet-emitting region. Composite disks retaining the outer solution confirm this numerically, requiring $\Mdot_{\rm out}=3.1$--$4.4\,\Msun\,\mathrm{yr^{-1}}$ and placing $\Rsph=(1.2$--$1.8)\times10^{3}\,\rg$ inside $\Ruv=(2.6$--$2.9)\times10^{3}\,\rg$. Replacing the inner $\sim10^{3}\rg$ by a supercritical flow shifts the fitted slope by only $\Delta\betaUV=+0.08$ to $+0.11$, comparable to uncertainties from standard spectral modelling. Reversing the radial ordering requires the effective transition to move outward by a factor $1.7$--$2.1$. Within our composite models, the measured slope constrains black-hole mass only to around $10^{6}\,\Msun$, an order of magnitude below the continuum Eddington mass. Continuum Eddington masses should therefore be treated as model-dependent bounds that assume global thin-disk efficiency. The same hierarchy appears in three additional JWST sources under the disk-dominated interpretation, where separation masses are only $4$--$6$\% of continuum Eddington masses. This distinction can substantially weaken the seed-mass and early-growth demands inferred from ultraviolet continua at cosmic dawn for the first massive black holes.
\end{abstract}


\section{Introduction}
\label{sec:intro}

The James Webb Space Telescope (JWST) has revealed accreting black holes throughout the first billion years of cosmic time, at masses and space densities that were not anticipated before launch \citep{Harikane2023, Maiolino2024JADES, Matthee2024, Greene2024, Kocevski2025}. Many are overmassive relative to their hosts, sharpening a long-standing question about their origin. They may descend from light Population~III remnants growing at or above the Eddington rate, or from heavy seeds formed by direct collapse or by runaway collisions in dense clusters \citep{MadauRees2001,BrommLoeb2003,Begelman2006,Volonteri2010,Inayoshi2020,ReganVolonteri2024}. Since the two channels differ by orders of magnitude in initial mass, the sharpest leverage comes from the highest redshifts, where the available growth time is shortest and an inferred black-hole mass translates most directly into a demand on the seed. GN-z11 at $z=10.603$ is the most distant such system currently known. First identified with HST \citep{Oesch2016} and confirmed spectroscopically with NIRSpec \citep{Bunker2023}, it is compact and unusually nitrogen-enhanced \citep{Tacchella2023,Cameron2023}, and \citet{Maiolino2024} argued for an AGN based on high-density ultraviolet lines, a fast C~\textsc{iv} outflow, and an extended Ly$\alpha$ halo \citep{Scholtz2024}.

Two independent interpretations of this source give black-hole masses differing by nearly an order of magnitude. A broad-line reading of N~\textsc{iv}]$\lambda1486$ gives $\log(\Mbh/\Msun)=6.2\pm0.3$ at a luminosity of order $5\Ledd$ \citep{Maiolino2024}, whereas fitting the standard-looking ultraviolet continuum with a radiatively efficient disk gives an Eddington mass $M_{\rm E}=1.12\times10^{7}\,\Msun$ \citep{Fabian2026}. The required growth differs correspondingly, the higher mass pointing to heavy seeds and the lower one to sustained supercritical accretion of the kind now invoked for several JWST populations \citep{Madau2014,Lupi2024,PacucciNarayan2024,Lambrides2026}. Ultra-deep spectroscopy from the Spectroscopic Ultra-deep Reionization-era Survey (SPURS) has since called both interpretations into question. The SPURS spectrum reveals P-Cygni stellar-wind features and broad He~\textsc{ii} emission, jointly reproduced by very massive stars at low metallicity, with the broad N~\textsc{iv}] component plausibly arising in dense WN or LBV-like winds, though an AGN contribution cannot be excluded \citep{Chen2026SPURS,NakaneOuchi2026}. With both the broad-line mass and the ultraviolet disk fraction now uncertain, it remains unclear what the continuum alone can require of the inner accretion flow.

The ultraviolet continuum slope illustrates the problem well. The measured slope of GN-z11 over $1400$--$3000\,\angstrom$, $\betaUV=-2.26\pm0.10$, lies close to the multitemperature-disk value of $-7/3$ \citep{Maiolino2024}, but it fixes the temperature profile only at the radii dominating that emission. Converting it into $M_{\rm E}$ further assumes this efficient solution extends inward to the innermost stable orbit, an assumption supercritical disks can violate, since they may be radially stratified, with a nearly Keplerian outer disk surrounding an interior regulated by advection, radiation transport, magnetic support, and mass loss \citep{ShakuraSunyaev1973,Abramowicz1988,Poutanen2007,Sadowski2014,Jiang2019,KubotaDone2019,Tang2019,Cheng2019}. \citet{Fabian2026} identified precisely such an interior as the lower-mass alternative, but left open whether its transition reaches the ultraviolet-emitting annuli strongly enough to alter the observed slope. In this manuscript, we separate what the continuum measures from the assumptions that follow only once a thin-disk solution is extrapolated inward. Section~\ref{sec:criterion} derives a closed-form criterion for where the supercritical transition lies relative to the observed continuum, and Section~\ref{sec:models} constructs composite disks that test it numerically. Section~\ref{sec:results} presents the resulting constraints on the ultraviolet slope and the black-hole mass, and Sections~\ref{sec:discussion} and \ref{sec:conclusions} discuss the consequences for continuum-based mass limits at cosmic dawn.

\section{Where the ultraviolet continuum is produced}
\label{sec:criterion}

Constraining the inner accretion state from the ultraviolet continuum requires a comparison of two radii: the radius dominating the observed emission, and the radius at which a supercritical flow departs from the radiatively efficient solution. Both are set by the same product $\Mbh\Mdot_{\rm out}$, so their ratio follows in closed form without specifying either factor individually. Far outside the inner boundary the effective temperature of a thin, radiatively efficient disk is \citep{ShakuraSunyaev1973,Pringle1981}
\begin{equation}
T_{\rm eff}(R)=\left(\frac{3G\Mbh\Mdot_{\rm out}}{8\pi\sigma R^3}\right)^{1/4},
\label{eq:teff}
\end{equation}
with $R$ the radius, $\Mdot_{\rm out}$ the outer mass-supply rate, $G$ the gravitational constant, and $\sigma$ the Stefan--Boltzmann constant. Electron scattering hardens the emergent spectrum, absorbed into a colour-correction factor $f_{\rm col}$ through the local intensity $B_\nu(f_{\rm col}T_{\rm eff})/f_{\rm col}^4$, a dilution preserving the bolometric flux \citep{ShimuraTakahara1995,Davis2005,Done2012}. For a flat, two-sided disk at inclination $i$, the isotropic-equivalent luminosity $L_\nu^{\rm iso}\equiv4\pi D^2F_\nu$ is
\begin{equation}
L_\nu^{\rm iso}=\frac{8\pi^2\cos i}{f_{\rm col}^4}\int B_\nu(f_{\rm col}T_{\rm eff})\,R\,dR ,
\label{eq:liso}
\end{equation}
where $B_\nu$ is the Planck function at frequency $\nu$, $D$ the luminosity distance, $F_\nu$ the observed flux density, and $\cos i$ the projection of the disk surface \citep{Mitsuda1984}.

Since $T_{\rm eff}\propto R^{-3/4}$, this integrand peaks sharply in $\ln R$ at $x_\nu\equiv h\nu/(kf_{\rm col}T_{\rm eff})=2.4325$, the root of $x=(8/3)(1-e^{-x})$. This defines the emission radius $\Rnu$, which dominates the observed continuum at $\nu$ and hence carries the measured slope \citep{DavisLaor2011}; we write $\Ruv$ for $\Rnu$ at $1400\,\angstrom$. The second scale is the spherization radius $\Rsph$, inside which the flux from one face exceeds $\Ledd/(4\pi R^2)$, radiation pressure drives mass loss, and Equation~(\ref{eq:teff}) fails \citep{ShakuraSunyaev1973,Poutanen2007}. Equating the two fluxes gives
\begin{equation}
\Rsph\simeq\frac{3\kappa\Mdot_{\rm out}}{8\pi c}=\frac{3}{2}\dot m_0\rg ,
\qquad
\dot m_0\equiv\frac{\Mdot_{\rm out}c^2}{\Ledd} ,
\label{eq:rsphapprox}
\end{equation}
with $\kappa$ the electron-scattering opacity and $c$ the speed of light. Using $\Ledd=4\pi G\Mbh c/\kappa$ and $\rg=G\Mbh/c^2$, the opacity cancels and $\dot m_0$ is the supply rate in units of $\Ledd/c^2$; lower-case radii denote $r\equiv R/\rg$ throughout.

Evaluating Equation~(\ref{eq:liso}) over $0<x<\infty$ and expressing the result via $\Rnu$ gives
\begin{equation}
\begin{split}
\nu L_\nu^{\rm iso}
&=\frac{60\,\Gamma(8/3)\,\zeta(8/3)\,x_\nu^{4/3}}{\pi^4}\cos i\,\frac{G\Mbh\Mdot_{\rm out}}{\Rnu}\\
&=3.894\,\cos i\,\frac{G\Mbh\Mdot_{\rm out}}{\Rnu},
\end{split}
\label{eq:power}
\end{equation}
with $\Gamma$ and $\zeta$ the gamma and Riemann zeta functions: the monochromatic luminosity measures the accretion power released interior to $\Rnu$. Since $\Rsph$ is proportional to the same product, eliminating $G\Mbh\Mdot_{\rm out}$ between Equations~(\ref{eq:rsphapprox}) and (\ref{eq:power}) yields
\begin{equation}
\frac{\Rsph}{\Rnu}=A_\nu\,\frac{\nu L_\nu^{\rm iso}}{\Ledd}\,\frac{1}{\cos i},
\label{eq:criterion}
\end{equation}
with $A_\nu=\pi^4/[40\,\Gamma(8/3)\,\zeta(8/3)\,x_\nu^{4/3}]=0.385$. Equation~(\ref{eq:criterion}) assumes the flat-disk $\cos i$ law rather than funnel or wind transfer \citep{Poutanen2007,Sadowski2014}, a radially constant $f_{\rm col}$, and the Newtonian asymptotic limit.

For GN-z11, with $\log(\nu L_\nu/\mathrm{erg\,s^{-1}})=44.3$ at $1400\,\angstrom$ \citep{Fabian2026}, $f_{\rm d}$ the disk fraction of that luminosity, the benchmark $\log(\Mbh/\Msun)=6.2$ \citep{Maiolino2024}, and $i=30^\circ$, Equation~(\ref{eq:criterion}) gives $\Rsph/\Ruv=0.44\,f_{\rm d}$, so that the separation mass defined by $\Rsph=\Ruv$ is
\begin{equation}
M_{\rm sep}=7.0\times10^{5}f_{\rm d}
\left(\frac{\nu L_{\nu,\rm obs}^{\rm iso}}{10^{44.3}\,\mathrm{erg\,s^{-1}}}\right)
\left(\frac{\cos i}{\cos30^\circ}\right)^{-1}\Msun .
\label{eq:msep}
\end{equation}
Taking $f_{\rm d}=1$ is the conservative choice, maximising the ratio; any stellar contribution only pushes the transition further inside the emitting region, from $0.31$ at the $70$ per cent disk fraction of \citet{Fabian2026} to smaller values still under the lower fractions favoured by recent ultra-deep spectroscopy \citep{NakaneOuchi2026,Chen2026SPURS}. The ordering $\Rsph<\Ruv$ is thus robust to the one decomposition parameter the continuum does not fix. But $M_{\rm sep}$ is only a geometric scale: it marks where inner-flow changes reach the emitting annuli, not how large a spectral signature they produce.

\section{Composite disks that isolate the inner flow}
\label{sec:models}

\begin{figure*}
\centering
\includegraphics[width=\textwidth]{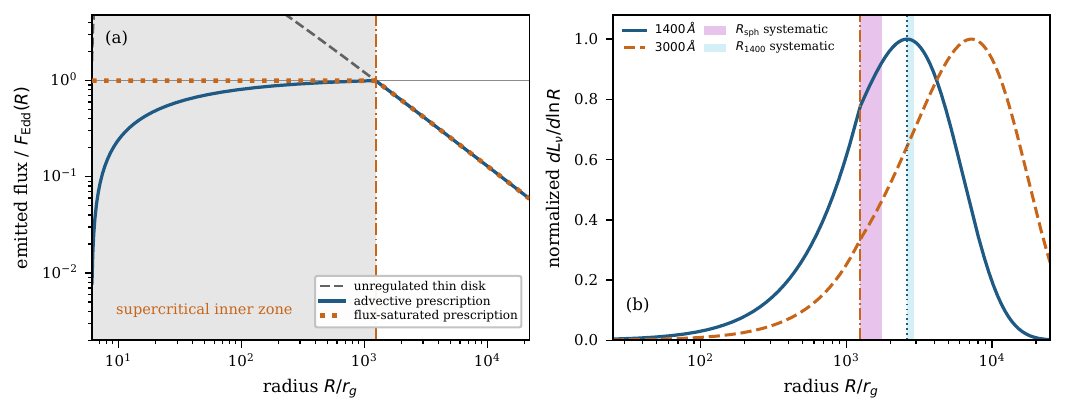}
\caption{Radial structure at the central N~\textsc{iv}-based benchmark mass for the fiducial normalization is shown. \textit{Left:} Emitted flux in units of the classical spherical-Eddington scale $\Ledd/(4\pi R^2)$ is shown as a function of radius. Gray dashed is the unregulated thin disk; blue solid and thick orange dotted are the advective and flux-saturated closures of Equations~(\ref{eq:fadv}) and (\ref{eq:fsat}). The orange dot-dashed vertical marks $\Rsph=1.23\times10^{3}\,\rg$. \textit{Right:} Normalized $dL_\nu/d\ln R$ at $1400\,\angstrom$ and $3000\,\angstrom$, is shown with vertical lines marking $\Rsph$ and $\Ruv$ and magenta and cyan bands their normalization systematics. Both contributions peak outside $\Rsph$, so the observed band is weighted toward radii where the standard outer profile survives.}
\label{fig:radial}
\end{figure*}

Isolating the effect of the inner flow requires a formalism that leaves the observed outer disk untouched and modifies only the interior, so that any resulting spectral change can be attributed to the inner flow alone. We therefore retain the zero-torque boundary term and adopt the one-sided outer-disk flux as given by:
\begin{equation}
F_{\rm thin}(R)=\frac{3G\Mbh\Mdot_{\rm out}}{8\pi R^3}
\left(1-\sqrt{\frac{R_{\rm in}}{R}}\right),
\label{eq:fthin}
\end{equation}
with $F_{\rm thin}$ the flux radiated per unit area by one face and $R_{\rm in}=6\rg$ the inner disk radius. Including this term, the outer root of the local-Eddington condition is \citep{Abolmasov2015}
\begin{equation}
r_{\rm sph}=2\dot m_0\cos^2\!\left[
\frac{1}{3}\cos^{-1}\!\left(-\frac{3\sqrt{r_{\rm in}}}{\sqrt{2\dot m_0}}\right)
\right],
\label{eq:rsph}
\end{equation}
which admits no supercritical annulus for $\dot m_0\leq4.5\,r_{\rm in}$, or $\dot m_0\leq27$ at $r_{\rm in}=6$. The interior flux, left unspecified by Equation~(\ref{eq:rsph}), depends on advection, wind work, vertical transport, magnetic support, and stress. Rather than solve the region, we bracket it with two literature-motivated closures: an advection-regulated profile
\begin{equation}
F_{\rm adv}(R)=F_{\rm thin}(R)\frac{R}{\Rsph},
\qquad R<\Rsph,
\label{eq:fadv}
\end{equation}
and a flux-saturated profile normalized to the classical spherical-Eddington scale,
\begin{equation}
F_{\rm sat}(R)=\frac{\Ledd}{4\pi R^2},
\qquad R<\Rsph.
\label{eq:fsat}
\end{equation}
Both join Equation~(\ref{eq:fthin}) continuously at $\Rsph$ and give $T_{\rm eff}\propto R^{-1/2}$ inside it. Their ratio $F_{\rm adv}/F_{\rm sat}=(1-\sqrt{R_{\rm in}/R})/(1-\sqrt{R_{\rm in}/\Rsph})$ departs from unity only in the boundary layer, reaching $0.99$ by $10^3\rg$, so we treat them as a single baseline case.

The outer supply is then fixed by the observed normalization through $\mathcal K\equiv\Mbh\Mdot_{\rm out}$, since $L_\nu\propto\mathcal K^{2/3}$ asymptotically. We carry two values  $\mathcal K=4.94\times10^6$ and $6.96\times10^6\,\Msun^2\,\mathrm{yr^{-1}}$ \citet{Fabian2026}, corresponding to their tabulated Eddington point and their direct $10^7\,\Msun$ \textsc{kerrbb} fit \citep{Li2005}. The $10^6\,\Msun$ fit of \citet{Fabian2026}, of statistically indistinguishable quality, gives $\mathcal K=6.2\times10^6\,\Msun^2\,\mathrm{yr^{-1}}$, which lies within this range and confirms $\mathcal K=(4.94$--$6.96)\times10^6\,\Msun^2\,\mathrm{yr^{-1}}$ as the normalization systematic carried throughout.

With $\mathcal K$ fixed, the full spectrum follows from integrating the composite flux profile across the disk, giving the angle-integrated, two-sided luminosity as: 
\begin{equation}
L_\nu=\frac{4\pi^2}{f_{\rm col}^4}
\int_{R_{\rm in}}^{R_{\rm out}}
B_\nu(f_{\rm col}T_{\rm eff})R\,dR,
\label{eq:spectrum}
\end{equation}
with $R_{\rm out}=3\times10^7\rg$ and $f_{\rm col}=1.7$ matching the published disk fits, and differing from Equation~(\ref{eq:liso}) by the factor $2\cos i$ converting an angle-averaged luminosity into the isotropic-equivalent value quoted by observers. At the fiducial normalization and benchmark mass this reproduces the measurement, returning $\log(\nu L_\nu^{\rm iso}/\mathrm{erg\,s^{-1}})=44.30$ at $1400\,\angstrom$; the high normalization overshoots by $23$ per cent ($44.39$), which is why we treat the fiducial case as primary. Every quoted $\betaUV$ is an unweighted least-squares fit in $\log F_\lambda$ versus $\log\lambda$ over $84$ logarithmically spaced wavelengths from $1400$--$3000\,\angstrom$. Because the dominant radii are $\gtrsim10^3\rg$, relativistic transfer and spin barely affect the fitted slope, though both remain important for the unobserved extreme ultraviolet and for $M_{\rm E}$.

\section{Results} \label{sec:results}
The analytical formalism of Section~\ref{sec:models}, applied at the GN-z11 mass estimation, fixes a definite chain of scales: the observed ultraviolet normalization sets the outer supply rate, the supply rate sets the transition radius, and comparison of that radius with the radii dominating the observed emission determines the resulting spectral response. We report this chain in order, and conclude by quantifying the mass constraint the ultraviolet slope alone supports.

\subsection{The low-mass branch is necessarily supercritical}
\label{sec:supercritical}

We first ask what the observed continuum demands of the outer supply rate at the N~\textsc{iv}]-based observed mass, before considering where the resulting transition lies. At $\Mbh=10^{6.2}\,\Msun$, the two continuum normalizations of Section~\ref{sec:models} require $\Mdot_{\rm out}=3.11$--$4.39\,\Msun\,\mathrm{yr^{-1}}$, corresponding to $\dot m_0=(8.8-12.5)\times10^{2}$ and $\Mdot_{\rm out}/\Mdot_{\rm Edd}=50$--$71$ for a radiative efficiency $\eta=0.057$. These values exceed the threshold $\dot m_0>27$ set by Equation~(\ref{eq:rsph}) by a factor of $33$--$46$, so a supercritical annulus is not marginally present but robustly established across the full normalization systematic; the resulting flux profile is shown in Figure~\ref{fig:radial}. If N~\textsc{iv} traces a broad-line region and a disk supplies the relevant ultraviolet component, the low-mass branch is therefore not merely compatible with supercritical accretion but requires it. Whether this supercritical interior is in practice detectable by JWST depends not on the existence of the transition but on its location relative to the ultraviolet-emitting region, which we examine next.

\begin{figure*}[t]
\centering
\includegraphics[width=\textwidth]{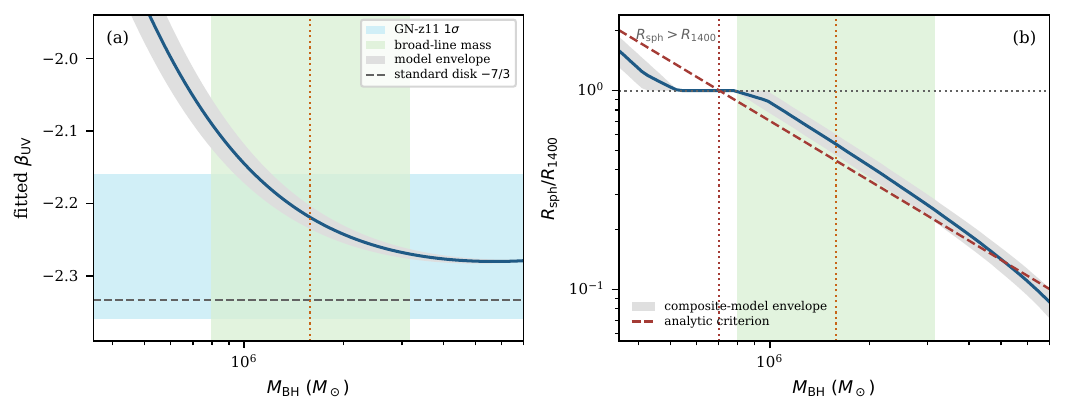}
\caption{\textit{Left:} Mass dependence of the two GN-z11 analysis, showing that they carry very different information. Gray envelopes span both continuum normalizations and both baseline inner closures;
green marks the $0.3$ dex N~\textsc{iv}]-based benchmark range and orange dotted its central value. Fitted $1400$--$3000\,\angstrom$ slope $\betaUV$, with the observed $1\sigma$ interval in cyan and
the standard-disk value $-7/3$ gray dashed. \textit{Right:} The ratio $\Rsph/\Ruv$ as a function of black-hole mass, with the analytic criterion of Equation~(\ref{eq:criterion}) shown as the red dashed curve and the separation mass $M_{\rm sep}$ marked by the red dotted vertical line. Unlike the fitted slope, this radial ratio responds cleanly and nearly inversely to mass, and the composite envelope tracks the analytic criterion closely wherever the two radii remain distinct.}
\label{fig:mass}
\end{figure*}

\subsection{The transition lies inside the ultraviolet-emitting region}
\label{sec:radial}

Having shown that the outer supply demands a supercritical interior, we now locate where that interior transitions to the standard thin-disk solution, and ask whether this transition reaches the radii producing the observed continuum. The composite profiles place it at $\Rsph=(1.23$--$1.76)\times10^{3}\rg$, while the $1400\,\angstrom$ emission peaks at $\Ruv=(2.60$--$2.93)\times10^{3}\rg$, so that $\Rsph/\Ruv=0.47-0.60$, below unity in every baseline model. The numerical separation mass, $6.2$--$7.0\times10^{5}\,\Msun$, reproduces the analytic value of Equation~(\ref{eq:msep}) to better than $12$\%, confirming that the asymptotic criterion of Section~\ref{sec:criterion} survives both the zero-torque boundary term and the full radial integration of Section~\ref{sec:models}.

The interior is nonetheless drastically restructured relative to a thin disk: the unregulated thin-disk flux rises above the local Eddington scale throughout the supercritical zone, which is what forces a regulated interior in the first place, and both closures cap the flux near that scale, departing from one another only in the innermost decade where the zero-torque factor suppresses $F_{\rm adv}$ relative to $F_{\rm sat}$ (Fig.~\ref{fig:radial}a). This restructuring is nevertheless poorly exposed by the observed band. The $1400\,\angstrom$ contribution to $dL_\nu/d\ln R$ peaks well outside $\Rsph$, and the $3000\,\angstrom$ contribution peaks farther out still, at $(7.2$--$8.1)\times10^{3}\rg$, nearly three times $\Ruv$ (Fig.~\ref{fig:radial}b), though $28$--$35$ per cent of the $1400\,\angstrom$ emission still arises inside the transition. The measured band is thus weighted toward radii where the standard outer solution survives, and reversing this ordering would require the effective transition to move outward by $\Ruv/\Rsph\simeq1.7$--$2.1$, the quantitative target any alternative inner solution must meet to become detectable in the present data.

\begin{figure*}
\centering
\includegraphics[width=\textwidth]{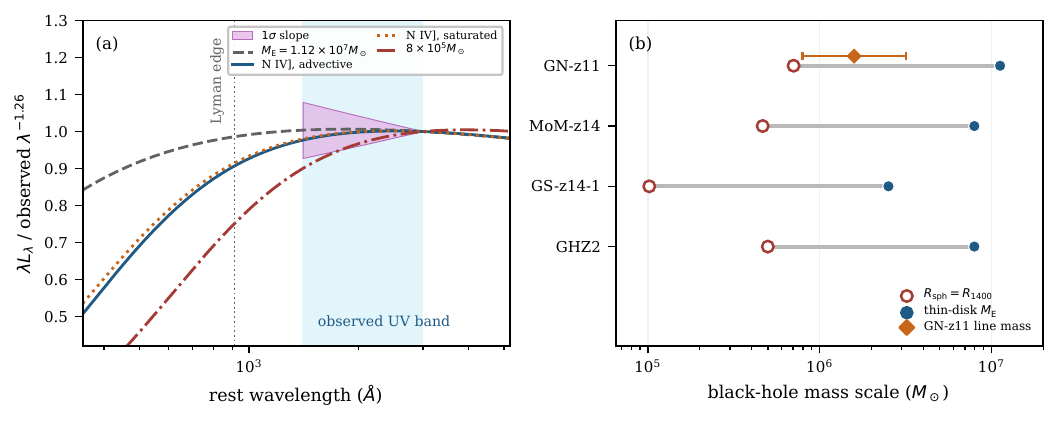}
\caption{\textit{Left:} Intrinsic thermal continua divided by the observed $\lambda L_\lambda\propto\lambda^{-1.26}$ shape and normalized at $3000\,\angstrom$, so that a perfect slope match is flat at unity. Magenta is the measured $1\sigma$ slope range, cyan is the observed band, and grey dotted is the Lyman edge. The gray dashed curve is the globally thin $M_{\rm E}=1.12\times10^{7}\,\Msun$ solution. At the N~\textsc{iv}]-based benchmark mass, $\Mbh=10^{6.2}\,\Msun$, the blue solid and orange dotted curves show the advective and flux-saturated closures evaluated with the same fiducial outer normalization; their difference therefore isolates the inner-flux prescription. The red dot-dashed curve shows an $8\times10^{5}\,\Msun$ model near $M_{\rm sep}$, also at the fiducial normalization, to illustrate the spectral curvature as the transition approaches the ultraviolet-emitting region. \textit{Right:} Separation masses from Equation~(\ref{eq:msep}) at $i=30^\circ$ are compared with the thin-disk Eddington masses $M_{\rm E}$ of \citet{Fabian2026}. The ultraviolet luminosities and slopes entering the comparison are from \citet{Maiolino2024}, \citet{Naidu2025}, \citet{Wu2025}, and \citet{Castellano2024} for GN-z11, MoM-z14, GS-z14-1, and GHZ2, respectively; the open circles are the $M_{\rm sep}$ values calculated here from the same $1400\,\angstrom$ luminosities. The orange diamond marks the conditional GN-z11 N~\textsc{iv}]-based benchmark. The three comparisons other than GN-z11 are illustrative and conditional on disk-dominated ultraviolet emission.}
\label{fig:sources}
\end{figure*}

\subsection{The ultraviolet slope cannot distinguish the inner states}
\label{sec:slope}

The radial separation established above suggests that the fitted slope should be largely insensitive to the inner solution, and the composite spectra confirm this: although the interior flux changes by more than an order of magnitude, the fitted slope shifts only marginally, from $\betaUV=-2.31$ for thin disks at the same mass ($-2.27$ at $M_{\rm E}$) to $-2.235$ to $-2.203$ for the composite models, all consistent with the observed $-2.26\pm0.10$ and therefore unable to distinguish among them. The relevant quantity is instead the response at fixed outer normalization,
\begin{equation}
\Delta\betaUV=\betaUV^{\rm composite}-\betaUV^{\rm thin}=+0.08\text{ to }+0.11 ,
\label{eq:dbeta}
\end{equation}
so that replacing the thin-disk solution across the entire inner $\sim10^{3}\rg$ shifts the slope by only $0.8$--$1.1$ times its present $1\sigma$ uncertainty. Figure~\ref{fig:sources}a makes this comparison at fixed outer normalization: the advective and flux-saturated spectra at the N~\textsc{iv}]-based benchmark mass nearly overlap throughout $1400$--$3000\,\angstrom$, showing directly that the measured band is insensitive to which baseline inner closure is adopted. The globally thin $M_{\rm E}$ solution is similarly close after normalization at $3000\,\angstrom$, while the $8\times10^{5}\,\Msun$ model near $M_{\rm sep}$ develops stronger curvature as the transition approaches the emitting region. Divided by the observed $\lambda L_\lambda\propto\lambda^{-1.26}$ shape, these branches remain within a few per cent of one another across the measured band and separate mainly shortward of it, toward the Lyman edge. Their extreme-ultraviolet behaviour is shown only illustratively, since photon trapping, disk-atmosphere structure, wind transfer, and intergalactic absorption are not modelled self-consistently there \citep{Ohsuga2003,Madau2026}.

This insensitivity survives substantially larger perturbations to the inner flux. Setting $F=\xi F_{\rm Edd,sph}$ with $\xi=0.3$ or $3$ moves $\betaUV$ only from $-2.12$ to $-2.36$, spanning the full observed $1\sigma$ interval without changing $\Rsph$. Varying the colour correction over $f_{\rm col}=1.4$--$2.4$ gives $-2.27$ to $-2.16$, and alternate fitting windows give $-2.25\lesssim\betaUV\lesssim-2.19$. Atmosphere and continuum-decomposition uncertainties therefore enter at the same level as an order-of-magnitude change in the inner flux itself, which is the practical statement of the degeneracy.

The bolometric output is not similarly protected. Integrating the composite profiles gives $L_{\rm bol}=4.8-6.7\,\Ledd$, and $\eta_{\rm eff}\equiv\frac{L_{\rm bol}}{\Mdot_{\rm out}c^2}=0.004-0.007$, consistent with the $\sim5\Ledd$ inferred at the benchmark mass \citep{Maiolino2024,Netzer2019}, though not independently, since that comparison relies on a $1400\,\angstrom$ bolometric correction. Over the same $\xi$ tests that move $\betaUV$ by $0.24$ dex, $L_{\rm bol}$ spans $2.6$--$17\,\Ledd$, a factor of $6.5$: the two observables respond to the inner closure on markedly different scales. Because $\eta_{\rm eff}$ falls an order of magnitude below the thin-disk value, a supply of tens of $\Mdot_{\rm Edd}$ can coexist with only a few $\Ledd$ of radiated luminosity while the outer ultraviolet continuum remains close to standard.

\subsection{What the ultraviolet slope does constrain}
\label{sec:massdep}

Having shown that the slope cannot separate the inner states at fixed mass, we have shown the mass information it does carry. The dependence is not flat everywhere: the slope flattens toward low mass, where the higher supply rate pushes $\Rsph$ into the emitting region, and steepens toward higher mass, saturating near $-2.28$ above $\sim3\times10^{6}\,\Msun$, where the transition has retreated well inside it (Fig.~\ref{fig:mass}a). Across most of the $0.3$ dex benchmark range, however, the composite slope varies by less than the measurement error, so the observed value does not by itself select a mass. Its only real discriminating edge lies at the shallow end, where the model slope becomes inconsistent with the observed $1\sigma$ interval below $M_{\betaUV}=(0.94\text{--}1.21)\times10^{6}\,\Msun$, a bound that weakens further to $(0.60$--$1.60)\times10^{6}\,\Msun$ once the colour correction is allowed to vary over $f_{\rm col}=1.4$--$2.4$. The measured continuum slope alone requires only $\Mbh\gtrsim10^{6}\,\Msun$, an order of magnitude below $M_{\rm E}=1.12\times10^{7}\,\Msun$; the discrepancy between these two values is attributable entirely to the assumption of global thin-disk efficiency.

In contrast to the fitted slope, the radial ratio $\Rsph/\Ruv$ carries a clean mass dependence throughout the same range. It declines monotonically, and approximately inversely, with mass; the composite envelope tracks the analytic criterion of Equation~(\ref{eq:criterion}) wherever the two radii remain distinct, and saturates at unity below $M_{\rm sep}$ because the numerical peak in $dL_\nu/d\ln R$ locks onto the flux break at $\Rsph$ rather than continuing to trace it (Fig.~\ref{fig:mass}b). This asymmetry is important, as $\betaUV$, the observable conventionally used to set $M_{\rm E}$, stays nearly flat across the mass range, while $\Rsph/\Ruv$ responds to it cleanly despite not being directly measured. Therefore, constraining the inner accretion state requires access to $\Rsph$, not a more precise measurement of $\betaUV$.

\section{Discussion}
\label{sec:discussion}

A standard-looking ultraviolet continuum and a standard inner accretion flow are not the same observable, and GN-z11 makes the distinction concrete. The continuum measures the temperature structure at the radii where the detected photons happen to originate, while a continuum Eddington mass demands, in addition, that the same radiatively efficient solution persist all the way inward to the black hole. If a disk contributes substantially, the conservative $f_{\rm d}=1$ case gives $\Rsph/\Ruv=0.47$--$0.60$ and $\Delta\betaUV=0.08$--$0.11$, and even at an inclination as large as $i=60^\circ$ the ratio only rises to $\Rsph/\Ruv=0.77$. A smaller disk fraction pushes the separation further still, and if stars dominate the ultraviolet altogether, the thin-disk continuum mass no longer constrains $\Mbh$ in any useful sense. A standard continuum, in other words, can coexist quite comfortably with a strongly supercritical inner $\sim10^{3}\rg$. The recent ultra-deep spectroscopy leaves both the disk fraction and the broad-line origin of N~\textsc{iv}] conditional \citep{NakaneOuchi2026, Chen2026SPURS}, making this distinction timely.

Whether this ordering survives depends on the theoretical treatment of the transition itself. Magnetic support, turbulent transport, advection, irradiation, mass loss, and wind transfer can each broaden or displace the region where the thin-disk solution fails \citep{Sadowski2014,Jiang2014,SadowskiNarayan2016,Jiang2019,Madau2026}, so the spectra computed here should be read as sensitivity calculations rather than as unique radiation-MHD predictions. Overturning the radial ordering of Section~\ref{sec:radial} requires the effective transition to move outward by a factor of $1.7$--$2.1$ before it can overtake the $1400\,\angstrom$ emission peak, while relativistic corrections of order $20$ per cent leave that ordering intact \citep{Abolmasov2015}. Radiation-MHD calculations can therefore be tested against a specific quantitative threshold, rather than only a qualitative expectation.

The radial separation also has consequences for the inferred growth rate. The required $3$--$4\,\Msun\,\mathrm{yr^{-1}}$ is an outer supply rate, not a rate through the horizon, and the two are only loosely coupled once a supercritical zone intervenes: the effective efficiency falls to $\eta_{\rm eff}=0.004$--$0.007$, an order of magnitude below the thin-disk value. Mass loss inside $\Rsph$ can reduce the rate actually reaching the black hole by a further order of magnitude, so it is retention and wind loss, not the outer ultraviolet slope, that set the pace of early growth. Probing that hidden region directly will require broad rest-optical lines, ultraviolet ionisation analysis, wind kinematics, continuum curvature, or deeper X-ray limits \citep{Lambrides2026}.

This framework extends immediately beyond GN-z11, subject only to the same continuum decomposition. Equation~(\ref{eq:criterion}) applies whenever a multitemperature disk contributes appreciably to the ultraviolet continuum, with $f_{\rm d}$ entering explicitly and a stellar-dominated source carrying no continuum Eddington mass constraint at all. Applied to the three further high-redshift sources tabulated by \citet{Fabian2026}, and conditional on disk-dominated ultraviolet emission, the separation masses come out at $1.0\times10^{5}$--$5.0\times10^{5}\,\Msun$, only $4$--$6$ per cent of their continuum Eddington masses \citep{Naidu2025, Wu2025, Castellano2024}, with the same order-of-magnitude gap present in every case (Fig.~\ref{fig:sources}b). The claim is not that these systems share a common inner flow, which the present data cannot address, but that in each of them the radius probed by the continuum lies far outside the radius at which the assumed efficient solution would fail. Wherever this separation holds, a black-hole mass inferred from continuum efficiency alone is a statement about the outer disk, not a lower bound on $\Mbh$ itself.

\section{Conclusions \& summary}
\label{sec:conclusions}

A standard-looking ultraviolet continuum tells us far less about the inner accretion flow of GN-z11 than has been assumed. Eliminating the outer supply rate between the monochromatic disk luminosity and the spherization radius gives a simple, closed-form criterion, $\Rsph/\Rnu=0.385\,(\nu L_\nu^{\rm iso}/\Ledd)/\cos i$, and this alone places the supercritical transition at $0.47$--$0.60$ of the $1400\,\angstrom$ emission radius at the N~\textsc{iv}]-based benchmark mass. Even replacing the entire inner $\sim10^{3}\rg$ with a supercritical flow shifts the fitted $1400$--$3000\,\angstrom$ slope by only $\Delta\betaUV=+0.08$ to $+0.11$, no larger than the shift from the colour correction or the fitting window alone. The measured slope, by itself, asks for little more than $\Mbh\gtrsim10^{6}\,\Msun$, a full order of magnitude below the continuum Eddington mass $M_{\rm E}=1.12\times10^{7}\,\Msun$. That entire gap rests on a single assumption: that thin-disk efficiency extends all the way in to the horizon.

Continuum Eddington masses should accordingly be reported as model-dependent bounds that presuppose global thin-disk efficiency, not as observational lower limits on $\Mbh$. The distinction has direct consequences. Across high-redshift samples it determines how strongly the observations actually demand massive seeds or sustained near-Eddington growth, and the criterion derived here applies to any JWST candidate for which the continuum decomposition and viewing geometry can be recovered. Breaking the degeneracy will require observables that probe inside $\Rsph$ itself, not tighter measurements of $\betaUV$. Broad rest-optical lines, wind kinematics, and deeper X-ray limits offer the most direct routes, and radiation-MHD calculations now have a specific quantity to test against: whether the effective transition can plausibly move outward by a factor of $1.7$--$2.1$ before it registers in the observed band. Testing that threshold against deeper JWST spectroscopy and radiation-MHD simulation will determine how securely the present ultraviolet continuum can constrain the black-hole mass and inner accretion state of GN-z11 at $z=10.6$.

\section*{Acknowledgments}
R.S. and S.M. acknowledge support from the Department of Atomic Energy, Government of India, under project no. RTI4019. The authors acknowledges the use of Generative AI for text editing.

\bibliographystyle{aasjournal}
\bibliography{references}

\end{document}